\documentstyle[amsfonts,amssymb,epsfig,12pt]{article}

\makeatletter
\renewcommand \thesection {\@arabic\c@section}
\renewcommand\thesubsection   {\thesection.\@arabic\c@subsection}
\renewcommand\thesubsubsection{\thesubsection .\@arabic\c@subsubsection}
\renewcommand\theparagraph    {\thesubsubsection.\@arabic\c@paragraph}
\renewcommand\section{\@startsection {section}{1}{\z@}%
                                   {-3.5ex \@plus -1ex \@minus -.2ex}%
                                   {1.9ex \@plus.2ex}%
                                   {\normalfont\large\bfseries\centering}}
\renewcommand\subsection{\@startsection{subsection}{2}{\z@}%
                                     {-2ex\@plus -1ex \@minus -.2ex}%
                                     {1.2ex \@plus .2ex}%
                                    {\normalfont\normalsize\bfseries\centering}
}
\renewcommand\subsubsection{\@startsection{subsubsection}{3}{\z@}%
                                     {-2ex\@plus -1ex \@minus -.2ex}%
                                     {.5ex \@plus .2ex}%
                                     {\normalfont\normalsize\em}}
\renewcommand\paragraph{\@startsection{paragraph}{4}{\z@}%
                                    {3.25ex \@plus1ex \@minus.2ex}%
                                    {-1em}%
                                    {\normalfont\normalsize\em}}
\renewcommand\subparagraph{\@startsection{subparagraph}{5}{\parindent}%
                                       {3.25ex \@plus1ex \@minus .2ex}%
                                       {-1em}%
                                      {\normalfont\normalsize\em}}
\makeatother

\newcounter{subequation}
	\newenvironment{subequation}%
	{\addtocounter{equation}{-1}%
	\stepcounter{subequation}%
	\begin{equation}}%
	{\end{equation}%
}

\newcommand{\beq}{\begin{equation}}
\newcommand{\eeq}{\end{equation}}
\newcommand{\bseq}{\begin{subequation}}
\newcommand{\eseq}{\end{subequation}}
\newcommand{\bea}{\begin{eqnarray}}
\newcommand{\eea}{\end{eqnarray}}

\newcommand{\refeq}[1]{(\ref{#1})}

\newcommand{\dd}{{\mathrm d}}
\newcommand{\pd}{\partial}

\newcommand{\rhs}{{\rm right\ hand\ side\ }}

\newcommand{\rank}{{\rm rank\ }}

\newcommand{\kb}{k_{\rm B}}
\newcommand{\eps}{\epsilon}

\newcommand{\noin}{\noindent}

\newcommand{\cG}{{\cal G}}
\newcommand{\cN}{{\cal N}}
\newcommand{\cM}{{\cal M}}

\newcommand{\RR}{{\Bbb R}}
\newcommand{\SS}{{\Bbb S}}

\newcommand{\pr}{\prime}
\newcommand{\ppr}{{\prime\prime}}

\newcommand{\QED}{{\rm Q.E.D.}}

\begin{document}

\title{SYMMETRY RESULTS FOR FINITE-\\
	TEMPERATURE, RELATIVISTIC\\
	THOMAS-FERMI EQUATIONS$^*$}

\author{Michael K.-H. Kiessling\\ 
\textit{Department of Mathematics, Rutgers University}\\
\textit{110 Frelinghuysen Rd., Piscataway, N.J. 08854}}

\bigskip
\date{Original version: Dec. 06, 2000; final: Dec. 14, 2001.
To appear in: \textbf{Communications in Mathematical  Physics}}

\maketitle

\begin{abstract}
\noindent
	In the semi-classical limit the relativistic quantum mechanics 
of a stationary beam of counter-streaming (negatively charged) electrons and 
one species of positively charged ions is described by a nonlinear system of 
finite-temperature Thomas--Fermi equations.
	In the high temperature / low density limit these
Thomas--Fermi equations reduce to the (semi-)conformal system of
Bennett equations discussed earlier by Lebowitz and the author. 
	With the help of a sharp isoperimetric inequality it is 
shown that any hypothetical particle density function which is not 
radially symmetric about and decreasing away from the beam's axis 
would violate the virial theorem. 
	Hence, all beams have the symmetry of the circular cylinder.
\end{abstract} 
\vskip 1truecm

\smallskip
\hrule
\smallskip
\noindent
$^*$\textit{In celebration of the $70^{th}$ birthday of Joel L. Lebowitz.}

\smallskip
\noindent
\copyright{2001} The author. Reproduction of this  article, in its
entirety, for non-commercial purposes is permitted.

\newpage
                  
\section{INTRODUCTION}
	Modern books on charged-particle beams,
e.g.~\cite{humphries}, usually contain a chapter about the Bennett 
model~\cite{bennettPAPa}, but back in the early 50's when regular
research on charged-particle beams came into sharper focus, W.H. Bennett's 
pioneering pre-WWII paper~\cite{bennettPAPa} on the statistical mechanics 
of a relativistic, stationary particle beam had  been forgotten, apparently,
and so in 1953 Bennett sent out a reminder note~\cite{bennettPAPb}.
	For some reason or other, Bennett's note did not appear until
1955~\cite{bennettPAPb}, the very year when Joel L. Lebowitz was launching 
his stellar career~\cite{joelA} with center of gravity in stationary 
non-equilibrium statistical mechanics~\cite{joelB, joelC, joelD}.
	At that time, a single issue of {\it The Physical Review} was still
of a decent size and could be consumed from first to last page by an 
individual with huge scientific appetite such as Joel, and Bennett's 
note~\cite{bennettPAPb} did not pass unnoticed before Joel's hungry eyes.
	All this happened a few years before I was born, but when I came
to spend some postdoctoral time with Joel nearly 40 years later, 
several interesting questions raised by Bennett's work were still unanswered,
and so we began to answer some of these~\cite{kiejllPhPl}.
	One of the problems we had to leave open was that of the symmetry 
of a beam. 
	Following Bennett we only inquired into circular-cylindrically 
symmetric solutions. 
	While it is a natural conjecture that in the absence of
external fields an unbounded straight particle beam with finite electrical 
current through its cross-section necessarily posesses the symmetry of the 
circular cylinder, how to prove it is not quite so obvious.
	It is with great pleasure when in this paper I present a
rigorous proof to Joel.

	Fitting for the occasion, the proof of the cylindrical 
symmetry of the beam involves statistical mechanics in an essential way.
	Namely, it is shown that any hypothetical 
stationary beam with finite electrical current whose particle density
functions are not radially symmetric about and decreasing away from 
the beam's axis would violate the virial theorem for this many-particle 
system.
	This symmetry proof covers Bennett's strictly classical model 
as well as its semi-classical upgrade, i.e. a system of relativistic, 
finite-temperature Thomas--Fermi equations which in the high-temperature 
/ low-density limit reduce to the (semi-)conformal Bennett equations. 
	The proof is, however, restricted to a system of merely two 
equations because the coefficient matrix for the beam equations has rank 2.
	Our symmetry theorem therefore does not apply to beams that consist 
of the negatively charged electrons and more than one, differently positively 
charged ion species; but then again, our method of proof not only yields the cylindrical
symmetry of the beam, it also yields monotonic radial decrease of
the particle density functions. 
	Hence, it is conceivable that  monotonicity of the density
functions may  be violated in an electron / multi-ion species beam 
while cylindrical  symmetry might still hold --- yet to prove that 
would seem to require an entirely new argument. 

	Incidentally, our result also sheds some new light on the 
theory of white dwarfs~\cite{chandra}.
	These Earth-sized, expired stellar objects shine in bright white light 
because they are still incredibly hot compared to our Sun, yet they are 
relatively cold compared to their Fermi temperature and therefore
essentially in their quantum ground state.
	This justifies using zero-temperature Thomas--Fermi theory for
the description of their overall structure~\cite{chandra} --- a fortunate
happening, for finite-temperature Thomas--Fermi theory could not be 
used in three dimensions since it does not have solutions with finite mass. 
	Interestingly, the finite-temperature Thomas--Fermi equations of the 
two-dimensional caricature of such a white dwarf star should have solutions 
with finite mass, because the gravitational potential in two dimensions is 
sufficiently strongly confining for this purpose. 
	In any event, relatively little is known 
rigorously\footnote{More is known rigorously~\cite{chandra, liebyau} 
	for the locally neutral approximation of the model, 
	where the positive and negative charges are distributed 
	identically and Coulomb's law is discarded.
	In particular, radial symmetry of solutions for 
	this locally neutral model has been proven by energy minimization 
	through radial rearrangement~\cite{liebyau}. 
	We remark that due to the enormous ratio of the electrical and 
	gravitational coupling constants the locally neutral approximation 
	is expected to be an excellent approximation for a white dwarf;
	however, this is not generally the case for a 
	particle beam, where the ratio of electric and magnetic 
	coupling constants may be arbitrarily close to $1$.}
for such a gravitating plasma of negative electrons and positive nuclei 
(all species treated as fermions) in either two or three dimensions; see 
the discussion of this model by W.E. Thirring in the preface to the E.H. Lieb 
jubilee volume~\cite{liebJUBILEE}, where Thirring gives an amusing account of
the pitfalls associated with the fact 
that the Thomas--Fermi equations are the Euler--Lagrange 
equations for the saddle points of a variational functional.  
	When dealing with saddle points, existence and symmetry of 
solutions via minimization by radial decreasing 
rearrangement~\cite{almgrenlieb, brothersziemer, carlenloss, kawohl} 
is not an option, and neither is symmetry via uniqueness by
convexity~\cite{kawohl} of the functional.

	Now recall that by the Biot-Savart law the magnetic interactions 
of straight, parallel electrical current filaments are attractive, with 
a distance law that is identical to the Newtonian gravity law in two 
dimensions.
	From this it follows that the finite-temperature Thomas--Fermi 
beam equations are identical to the finite-temperature Thomas--Fermi 
equations of the two-dimensional caricature of a white dwarf model, 
with the magnetic flux function re-interpreted as the gravitational 
Newton potential in two dimensions, and the mean electric current of 
each species (positive after at most a joint space rotation) re-interpreted 
as the mass of that species. 
	Our symmetry result can be rephrased thus: two-dimensional 
finite-temperature white dwarfs are radially symmetric.
	
	Our proof of symmetry, which is based on the 
Rellich~\cite{rellich}--Pokhozaev~\cite{pokhozaev} identity 
(which expresses the virial theorem) and the classical 
isoperimetric inequality~\cite{bandleBOOK, brothersziemer},
does involve radial rearrangements in a  
strategy that goes back at least as far as~\cite{bandleBOOK}, 
where it is applied to Liouville's 
equation\footnote{The elliptic Liouville equation, known from 
		two-dimensional differential geometry, is meant
		and not the evolution equation on phase space 
		known from statistical mechanics.}
in a disk $\subset\RR^2$~\cite{bandleBOOK}. 
	In~\cite{chakieGAFA} this strategy was generalized to systems of 
PDEs of Liouville type in all $\RR^2$ which are unrestricted in size but 
which have a symmetric, fully stochastic coefficient matrix of full rank.
	The Bennett equations also constitute a  Liouville system, but 
are  not covered by the theorem of~\cite{chakieGAFA} because 
their coefficient matrix is generally not symmetric, has some 
negative elements, and is always rank 2. 
	The present paper develops the necessary generalizations 
of~\cite{chakieGAFA} to overcome the first two peculiarities of the 
Bennett equations, but the rank 2 restricts the proof to a system of
two equations.
	By adapting the treatment of single PDEs with more general 
nonlinearities developed in~\cite{lionsPAPa} (cf. also~\cite{kesavanpacella})
and~\cite{chakieCR, chakieNA} to the system case we are able to extend our 
proof of symmetry for the Bennett equations to the relativistic, 
finite-temperature Thomas--Fermi beam equations.  

	Our proof simplifies considerably when the systems of
Thomas--Fermi and Bennett equations are restricted to a disk 
with $0$-Dirichlet boundary conditions for the electric and 
magnetic potentials. 
	In this compact case, an alternate proof of the radial
symmetry and decrease of the solutions to systems of PDE which 
includes the finite-temperature Thomas--Fermi and Bennett equations, 
was given by Troy~\cite{troy}, who exploited  Alexandroff's method of
moving planes.
	For more on the moving-planes method, 
see~\cite{serrin, gidasninirbg, li, chenli, chakieCMP, chakieDUKE}.
	Troy's proof has been extended to Liouville systems in
unbounded domains, the Bennett equations not included though,
in~\cite{chipotshafrirwolansky}.
	Presumably, the moving planes method can be made to work 
also for the system of Thomas--Fermi equations studied here; however, 
this is not done in this paper. 

	While the present paper addresses only the question whether 
invariance of the PDEs under rotations implies radial symmetry of
their solutions, these PDEs feature other symmetries which deserve
mentioning.
	The system of Thomas--Fermi equations is invariant under 
the isometries of Euclidean space, simple gauges, and Lorentz boosts
along the beam.	
	The Bennett equations are in addition to that invariant under 
isotropic scaling in $\RR^2$, and  for a special family of parameter 
values also under Kelvin transformations, in which case 
they are invariant under the Euclidean conformal group of $\RR^2$.
	In this fully conformal case the conformal orbit of the finite current 
solutions is connected and itself invariant~\cite{kiejllPhPl}.
	Invariance under the Euclidean conformal group holds also for
the Liouville systems studied in~\cite{chakieGAFA}, but their conformal 
orbit of finite mass solutions is generally not connected, and each 
component not invariant under inversions~\cite{chipotshafrirwolansky}.
	Toda systems in $\RR^2$, which are Liouville systems with
symmetric coefficient matrix given by the $SU(N)$ Cartan 
matrix, are studied in~\cite{jostwangA, jostwangB}.
	The distribution of negative and positive signs in the $SU(2)$ 
Cartan matrix is opposite to that in our Bennett equations, and sure 
enough, our radial symmetry proof fails in this case.
	Interestingly, in this case one can show that radial symmetry is
in fact broken by some solutions, see the bifurcation argument 
with $n=2$ in (1.7) of~\cite{chakieGAFA}, and see~\cite{jostwangB} for 
the construction of the complete solution family with finite masses.
	Another interesting topic not discussed further here is 
whether the translation invariance along the beam can be broken, as 
is suggested by various dynamical beam instabilities~\cite{weinberg}.

	The remainder of this paper is structured as follows.
	In the next section we formulate the basic equations of
the semi-classical beam model and its classical limit.
	Existence of solutions is briefly touched upon.
	In section 3 we state our two main theorems, and in section 4
we present their proofs.


\section{RELATIVISTIC BEAM EQUATIONS}
	We let $a\in\SS^2$ denote the fixed axis of the beam, 
$x\in\RR^2$ a point in the cross-section of the beam containing 
the coordinate origin, and $p\in \RR^3$  the  kinematical particle 
momentum.
	The self-consistent electric field of the beam is given by
$E(x)=-\nabla\phi(x)$, where $\phi$ is the electric potential, 
and the magnetic field by $B(x)= \nabla \psi(x)\wedge a$, where
$\psi$ is the magnetic flux function.
	The beam consists of spin $1/2$ electrons (negatively charged, 
thus indexed by $s=-$) and one species of positively charged spin $1/2$ 
fermions (indexed by $s=+$),
characterized by the following parameters: the particle charges $q_s$ and rest 
masses $m_s$; the rest frame temperatures $T_s$; the external 
chemical potentials $\mu_s$; and lab frame drift speeds $c\nu_s$, 
where $c$ is the speed of light and $\nu_s\in (-1,1)$.
	We demand $\nu_+\neq \nu_-$, as appropriate 
for counter-streaming particle species. 
	The temperatures and drift speeds combine into the thermal 
lab frame parameters $\beta_s^{-1} = \kb T_s \sqrt{1 -\nu_s^2}$.

\subsection{The semi-classical model (Thomas--Fermi theory)}
	The finite-temperature Thomas--Fermi model of a straight,
relativistic beam is set up as follows. 
	In the lab frame the density of $s$-particles at $x$ is given by 
$\rho_s(x)={\cG}_s^{\rm TF}\!\left(\phi,^{\phantom{b}}\!\!\psi\right)(x)$, 
where
\beq
	{\cG}_s^{\rm TF}(\phi,\psi) 
= 
	\frac{2}{h^3}\int_{\RR^3}
\frac{\dd p}{ 1+e^{-\beta_s\! 
\left(\mu_s-c\sqrt{m_s^2c^2+|p|^2}+\nu_scp\cdot a-q_s[\phi-\nu_s\psi]\right)}}
\label{TFdensity}
\eeq
is the finite-temperature Thomas--Fermi density function for the 
relativistic $s$-species, which is subjected to the integrability condition 
\beq
\int_{\RR^2} 
{\cG}_s^{\rm TF}\!\left(\phi,^{\phantom{b}}\!\!\psi\right)(x)\dd x
=
N_s,
\label{Ns}
\eeq
where $N_s$ is the number of $s$-particles per unit length of beam.
	The phase-space density function under the integral in 
\refeq{TFdensity} is the drifting Fermi--Dirac--J\"uttner 
function~\cite{juettnerQM} with local chemical self-potential 
$-q_s(\phi(x)-\nu_s \psi(x))$.
	The electric charge and current densities in the Poisson equations 
for the electric potential $\phi$ and the magnetic flux function $\psi$
are computed with the density functions \refeq{TFdensity}, which  
leads to the  system of nonlinear PDEs
\bea
&&
  -\Delta \phi \,
= 
  4\pi{\textstyle\sum_s}\ q_s\, {\cG}_s^{\rm TF}\!\left(\phi, \psi\right)
\label{TFeqPHI}
\\
&&
  -\Delta \psi 
= 
 4\pi{\textstyle\sum_s} \nu_s q_s {\cG}_s^{\rm TF}\!\left(\phi, \psi\right) 
. \label{TFeqPSI}
\eea
	Here and in the following, $\sum_s$ or $\sum_t$ always stands for
summation over the particle species, i.e. $s=\mp$ and $t=\mp$. 

	The Thomas--Fermi equations \refeq{TFeqPHI}, \refeq{TFeqPSI}, are
invariant under the isometries of three-dimensional Euclidean space,  
Lorentz boosts along the beam's axis $a$,  and the gauge transformation
\beq
	\phi(x)\to\phi(x)+\phi_0;
\quad  \psi(x)\to\psi(x)+\psi_0;
\quad  \mu_s\to \mu_s +q_s(\phi_0-\nu_s\psi_0)
,\eeq 
where $\phi_0$ and $\psi_0$ are arbitrary constants.

	Since we are interested in the beam's natural symmetries,
we will not allow ``sources at infinity'' which would deform the 
beam; hence, we supplement \refeq{TFeqPHI} and \refeq{TFeqPSI}
with the asymptotic conditions that, uniformly as $|x|\to\infty$, 
\beq
\lim_{|x|\to \infty}\ \frac{\phi(x)}{ Q \ln \frac{1}{|x|}}
 \ = 2 =\ 
 \lim_{|x|\to \infty}\ \frac{c \psi(x)}{I \ln \frac{1}{|x|}} 
, \label{AC}
\eeq
with $I\neq 0$ and $Q\neq 0$, where $I=\sum_s N_sq_s\nu_sc$ is the total 
electrical current through the beam's cross-section and $Q=\sum_s N_sq_s$ 
the total charge per unit length of beam  in the lab system; 
if $Q=0$, the left equation in \refeq{AC} is to be replaced 
by the condition that $\phi(x)\to\, const$ uniformly as $|x|\to\infty$.
	The situation $I=0$ is not considered here, for then 
of course there is no stationary beam.

\smallskip\noin
{\it Remark:} There are good reasons to conjecture that the asymptotic 
conditions \refeq{AC} {are in fact implied by} 
\refeq{TFdensity}-\refeq{TFeqPSI}.
	Analogous results have been proven for Liouville's 
equation~\cite{chenli} and for some Liouville 
systems~\cite{chakieGAFA, chipotshafrirwolansky}.
	No attempt will be made here to  generalize
these results to \refeq{TFdensity}-\refeq{TFeqPSI}. 
	However, we note that such a generalization would have the 
interesting {\it physical} implication (within the limits of applicability 
of the model) that one cannot maintain a stationary straight beam of finite 
current, whatever the geometry of its cross section, when there are 
magnetic or electric multipole sources ``at infinity.'' 
\smallskip

	To the best of the author's knowledge, the existence of beam 
solutions in the Thomas--Fermi model \refeq{TFdensity}-\refeq{TFeqPSI}
with asymptotics \refeq{AC} has not yet been studied rigorously.
	However, this semi-classical model is surely more regular than
the classical one, addressed next.
\subsection{The classical limit (Bennett theory)}
	In the high-temperature / low-density limit, i.e. formally 
 $0<\beta_s\ll 1$ and $\beta_s\mu_s \ll -1$, 
the Fermi--Dirac--J\"uttner functions \cite{juettnerQM} reduce to  
the Maxwell--Boltzmann--J\"uttner functions \cite{juettnerCLASSa} 
(see also \cite{degrootBOOK},  p.46, eq. (24)), 
so that the Thomas--Fermi densities \refeq{TFdensity} simplify to 
Boltzmann densities,
\beq
	{\cG}_s^{\rm B}(\phi,\psi) 
= 
	\frac{2}{h^3}\int_{\RR^3} 
e^{-\beta_s\! \left(c\sqrt{m_s^2c^2+|p|^2} - \nu_s cp\cdot a \right)}
	\dd p \, e^{\beta_s (\mu_s - q_s[\phi -\nu_s\psi])} 
,\label{Bdensity}
\eeq
and \refeq{Ns} becomes 
\beq
\int_{\RR^2} 
{\cG}_s^{\rm B}\!\left(\phi,^{\phantom{b}}\!\!\psi\right)(x)\dd x
=
N_s.
\label{NsB}
\eeq
	The system of equations \refeq{TFeqPHI} and \refeq{TFeqPSI} 
then reduces to the Bennett equations
\bea
&&	
	-\Delta \phi 
= 
	4\pi {\textstyle\sum_s} N_s q_s 
   \frac{\, e^{-\beta_s q_s(\phi - \nu_s \psi)} }
	{\int_{\RR^2}e^{-\beta_s q_s(\phi-\nu_s \psi)}\dd x}
\label{BeqPHI}
\\
&&
	-\Delta \psi
= 
	4\pi  {\textstyle\sum_s} N_{s} q_s \nu_{s} 
   \frac{\,e^{-\beta_s q_{s} (\phi - \nu_{s} \psi)}}
	{\int_{\RR^2}e^{-\beta_s q_{s} (\phi - \nu_{s} \psi)} \dd x} 
, \label{BeqPSI}
\eea
see \cite{bennettPAPa} eq.'s(8),(9), and  \cite{bennettPAPb} 
eq.(7),\footnote{ 
		In his papers~\cite{bennettPAPa, bennettPAPb}, Bennett 
		employed a classical, semi-relativistic setup, assuming 
		drifting Maxwell-Boltzmann distributions with relativistic
		drift speeds, yet with 	non-relativistic velocity
		dispersion in the cross-section	of the beam; the 
		relativistic model with drifting J\"uttner functions
		was used in \cite{benford}. It should be noticed, though, 
		that after integration over momentum space the very system 
		of equations \refeq{BeqPHI}, \refeq{BeqPSI} results in either
		case, and it does so also in the strictly non-relativistic 
		limit~\cite{kiejllPhPl} --- except for minor re-interpretations
		of the parameters in each case.}
where we have eliminated the external chemical potentials $\mu_s$ via 
\refeq{NsB}.

	The Bennett system is invariant under the isometries of 
three-dimensional Euclidean space and under Lorentz boosts along 
the beam's axis, $a$.
	Restricted to the beam's cross-section, it is also invariant 
under isotropic scaling, and in the special case when the parameters satisfy
\beq
\beta_s q_s  (\nu_s c^{-1}I - Q) = 2,\qquad s = \mp
, \label{conform}
\eeq
also invariant under translated inversions. 
	Thus, \refeq{conform} implies invariance of the Bennett system
under the conformal group of two-dimensional Euclidean space, acting
in the beam's cross-section.
	In addition, the Bennett equations are invariant under
a gauge transformation $\phi(x) \to \phi (x)+\phi_0$, 
$\psi(x) \to \psi(x)+ \psi_0$.
	Recall that we already eliminated the external chemical potentials
via the constraint equations \refeq{Ns} in the Bennett limit.

	In the conformally invariant case \refeq{conform}, 
Bennett's Ansatz\footnote{Bennett actually made the Ansatz that 
		$\rho_+(x)/\rho_-(x) =\ const$, which up to  gauge 
		freedom for the potentials is equivalent to 
		\refeq{AgleichPHI}.}
\beq
	I^{-1}c\psi(x) =  v(x) = Q^{-1}\phi(x) 
\label{AgleichPHI}
\eeq
maps \refeq{BeqPHI} and \refeq{BeqPSI} separately into
Liouville's equation~\cite{liouville}
\beq
	-\Delta {v} 
= 
	4\pi \frac{\;e^{2v}}
		  {\int_{\RR^2}e^{2v}\,\dd x}
. \label{liouEQ}
\eeq
	As remarked above, it has been proven in \cite{chenli} that any 
regular solution of \refeq{liouEQ}, with the understanding that
$\int\exp(2v)\dd x <\infty$, satisfies 
\beq
	\lim_{|x|\to\infty}\ \frac{v(x)}{\ln\frac{1}{|x|}} 
= 
	 2
,\label{PSIasympCONF}
\eeq
uniformly as $|x|\to\infty$, which implies
that the asymptotic conditions \refeq{AC} are 
automatically satisfied if $\phi$ and $\psi$ are given by \refeq{AgleichPHI}.
	It has also been proven in \cite{chenli}, and subsequently
in \cite{chakieGAFA, chouwan} by using alternate techniques,
that \refeq{liouEQ} has only one regular family of solutions, given by
\beq
v(x|x_0;k) = v_0 + \ln \frac{1}{1+k^2 |x-x_0|^2} 
, \label{BpinchPOT}
\eeq
where $k^{-1}>0$ is an arbitrary scale length, $x_0$ the
arbitrary center of rotational symmetry of the solution, and
$v_0$ an arbitrary gauge constant.
	The corresponding current density $j(x)$ and charge density 
$q(x)$ are given by
\beq
	I^{-1} j(x) 
=  
	\frac{1}{\pi}\frac{k^2}{\left(1+ k^2 |x-x_0|^2\right)^2} 
= 
	Q^{-1}q(x)
. \label{BpinchDENS}
\eeq
	The density profile \refeq{BpinchDENS} is the celebrated
Bennett beam profile.

	Bennett speculated about the existence of other solutions to
\refeq{BeqPHI} and \refeq{BeqPSI} with asymptotics \refeq{AC};
see \cite{bennettPAPa} p.893, and \cite{bennettPAPb} p.1587. 
	(In the punctured plane additional solutions are readily found, 
see e.g. \cite{benford}; however, they all lack regularity, due
to a point source, at the origin.)
	In~\cite{kiejllPhPl} we proved that in the conformal case
\refeq{conform}, Bennett's system of equations
\refeq{BeqPHI} and \refeq{BeqPSI}, supplemented by the asymptotic conditions
\refeq{AC}, are in fact equivalent to \refeq{liouEQ} (with
asymptotic condition \refeq{PSIasympCONF} automatically satisfied,
see above) so that \refeq{BpinchPOT} then exhausts all possibilities. 
	Moreover, for the semi-conformal case where \refeq{conform} does 
not hold  we proved the existence of a continuous parameter family of 
smooth radial solutions to \refeq{BeqPHI} and \refeq{BeqPSI} with asymptotics 
\refeq{AC} which are not invariant under inversions. 

	All the solutions of our beam equations are 
automatically also stationary solutions of the equations of Vlasov's 
relativistic kinetic theory~\cite{vlasovBOOK}.
	In \cite{kiejllPhPl} we showed that the Bennett equations 
can also be realized as transversal part of stationary 
{\it dissipative} kinetic equations in which the dissipation, 
modeled by a thermostat, compensates the action of an applied 
longitudinal electromotive force that drives the current.	

	In \cite{kiejllPhPl} we also gave a rigorous proof that
all radial solutions of \refeq{BeqPHI} and \refeq{BeqPSI} satisfying
\refeq{AC} also satisfy the Bennett identity
\beq
	c^{-2} I^2 - Q^2  
= 
	2 {\textstyle\sum_s} N_s\kb T_s \sqrt{1 -\nu_s^2} 
. \label{bennettID}
\eeq 
	The identity \refeq{bennettID} was originally obtained by
Bennett~\cite{bennettPAPb} in a formal (and not entirely compelling) 
manner by studying the radial time-dependent virial.
	In this paper we will show that the Bennett identity 
\refeq{bennettID}, respectively it's counterpart for the Thomas--Fermi 
model, holds {\it a priori} without assuming symmetry, and this fact will
be one major ingredient in our proof of the cylindrical symmetry of
the beams.

\section{MAIN RESULTS} 
	To state our virial theorem, we introduce the thermodynamic
potentials (per unit length of beam), given by
\beq
J^{\rm TF}\!
= 
\!\sum_s\beta_s^{-1}\frac{2}{h^3}\int_{\RR^2}\!\int_{\RR^3}\!
\ln\left[
 1+e^{\beta_s\! \left(
	\mu_s - c\sqrt{m_s^2c^2+|p|^2} +\nu_s cp\cdot a -q_s[\phi -\nu_s\psi] 
			\right)}\!
	\right]\!\dd p\;\dd x
\label{THERMpotTF}
\eeq
for the semi-classical model, respectively by
\beq
J^{\rm B}
= 
\sum_s\beta_s^{-1}\frac{2}{h^3}\int_{\RR^2}\int_{\RR^3}
e^{\beta_s\! \left(
	\mu_s - c\sqrt{m_s^2c^2+|p|^2} +\nu_s cp\cdot a -q_s[\phi -\nu_s\psi] 
			\right)}
\dd p\;\dd x
\label{THERMpotB}
\eeq
for the classical model.
\smallskip

{\bf Theorem 3.1: } {(Virial identity.) \it 
	Let $\phi \in C^{2,\alpha}(\RR^2)$ and $\psi\in C^{2,\alpha}(\RR^2)$ 
solve \refeq{TFeqPHI} and \refeq{TFeqPSI} under the constraints
\refeq{Ns}, respectively solve \refeq{BeqPHI} and \refeq{BeqPSI} 
under the constraints \refeq{NsB}, $s=\mp$, in either case
subjected to the asymptotic conditions \refeq{AC}.
	Then 
\beq
	c^{-2} I^2 - Q^2  
= 
	2 J
, \label{virialID}
\eeq 
where $J$ stands for either $J^{\rm TF}$ or $J^{\rm B}$. }
\newpage


	We also show that deviations from cylindrical symmetry
violate \refeq{virialID}, which is expressed in the next theorem.
\smallskip

{\bf Theorem 3.2: } {(Cylindrical symmetry.) \it 
	Let $\phi \in C^{2,\alpha}(\RR^2)$ and $\psi\in C^{2,\alpha}(\RR^2)$ 
solve \refeq{TFeqPHI} and \refeq{TFeqPSI} under the constraints
\refeq{Ns}, respectively solve \refeq{BeqPHI} and \refeq{BeqPSI} 
under the constraints \refeq{NsB}, $s=\mp$, 
subjected to the asymptotic conditions \refeq{AC}.
	Then there exists a point $x_0\in\RR^2$ such that both
$\phi$ and $\psi$ are  radially symmetric about $x_0$, and the density 
functions ${\cG}_s\!\left(\phi,^{\phantom{b}}\!\! \psi\right)(x)$ 
are decreasing away from $x_0$, where ${\cG}_s$ here stands for
either the Thomas--Fermi or the Boltzmann density function.}


\section{PROOFS}
	We rewrite the Thomas--Fermi, respectively
Bennett system in two equivalent versions, which may be called the 
``density potential representation'' and 
the ``chemical self-potential representation.'' 
	We will switch between these representations at our
convenience to obtain the asymptotic estimates, as $|x|\to\infty$, and the 
isoperimetric estimates needed for our proofs of Theorems 3.1 and 3.2.
	
\subsection{The alternate PDE representations}
	The chemical self-potentials $U_s(x),\ x \in \RR^2$, are given by
\beq
	U_s = - q_s( \phi - \nu_s \psi) 
. \label{chemUs}
\eeq
	We also introduce density potentials $u_s(x),\ x \in \RR^2$, 
defined by the  invertible linear system
\bea
&& \phi = {\textstyle\sum_s}  q_s u_s ,
\label{PHIus}
\\
&& \psi = {\textstyle\sum_s}  \nu_s q_s u_s 
.\label{PSIus}
\eea
	Clearly, 
\beq
U_s =  {\textstyle\sum_t}\gamma_{s,t} u_{t}
,\label{usUs}
\eeq
where 
\beq
\gamma_{s,t} =  - q_s q_t  (1 - \nu_s \nu_t)
\label{gMATRIX}
\eeq
denotes the entries of the matrix of coupling constants.
	Notice that
\beq
\det (\gamma) = - (q_+ q_-)^2 (\nu_+ - \nu_-)^2
, \label{detMATRIX}
\eeq
so that for $\nu_+\neq \nu_-$, we have 
\beq
\rank (\gamma) = 2
,\label{rankMATRIX}
\eeq
hence
\beq
u_s = {\textstyle\sum_t}\gamma_{s,t}^{-1} U_{t}
\label{Usus}
\eeq
where $\gamma_{s,t}^{-1}$ denotes the entries of the inverse
matrix $\gamma^{-1}$ to $\gamma$.

	Now let ${\cG}_s$ stand for either ${\cG}_s^{\rm TF}$ 
or ${\cG}_s^{\rm B}$.
	We note that ${\cG}_s (\phi,\psi)$ depends on $\phi$ and
$\psi$ only through the combination $-q_s(\phi -\nu_s \psi)=U_s$; thus 
we can write ${\cG}_s(\phi,\psi)= G_s(U_s)=G_s(\sum_t\gamma_{s,t}u_t)$, 
where of course $G_s$ stands for either $G_s^{\rm TF}$ or $G_s^{\rm B}$.
	In either case, the map $w\mapsto G_s(w)$ is monotonic
increasing.

	It then follows at once that the chemical self-potentials $U_s$
solve
the system of nonlinear PDEs
\beq
-\Delta U_s = 4\pi {\textstyle\sum_t}\gamma_{s,t}
G_{t}( U_{t}) 
,\label{UsPDE}
\eeq
supplemented by the integrability conditions
\beq
	\int_{\RR^2} G_s(U_s)\dd x 
= 
	N_s 
\label{UsNs}
\eeq
and by the asymptotic conditions that, uniformly as $|x|\to\infty$,
\beq
	\lim_{|x|\to\infty}\ \frac{U_s(x)}{\ln\frac{1}{|x|}} 
= 
	 2 {\textstyle\sum_t}\gamma_{s,t}N_t
.\label{UsAC}
\eeq

	Alternately, in terms of the $u_s$ 
we get the following representation for our Thomas--Fermi / Bennett models,
\beq
-\Delta u_s = 4\pi G_s\big( {\textstyle\sum_t}\gamma_{s,t}u_t\big) ,
\label{usPDE}
\eeq
supplemented by the integrability conditions
\beq
\int_{\RR^2} G_s\big( {\textstyle\sum_t}\gamma_{s,t}u_t\big)\;\dd x 
= 
	N_s 
\label{usNs}
\eeq
and by the asymptotic conditions that, uniformly as $|x|\to\infty$,
\beq
	\lim_{|x|\to\infty}\ \frac{u_s(x)}{\ln\frac{1}{|x|}} 
= 
	 2N_s,
\label{usAC}
\eeq
for $s=\mp$. 
	This constitutes the density potential representation of our
Thomas--Fermi / Bennett models.
\smallskip

\noin
{\it Remark:}
	For the sake of completeness, we also state the  PDEs of
the Bennett model explicitly as a Liouville system. 
	We readily eliminate the $\mu_s$ in terms of the $N_s$, using
\refeq{usNs}. 
	Setting now $u_s = N_s v_s$ and $\beta_s\gamma_{s,t}N_t =\kappa_{s,t}$,
and furthermore ${\textstyle\sum_t}\kappa_{s,t}v_t = 2V_s$ 
(equivalently, $\beta_s U_s= 2V_s$), with $s$ and $t$ taking the 
``values'' $\pm$, we rewrite \refeq{usPDE} into the form
\beq
-\Delta v_s = 4\pi 
   \frac{\, \exp\left( {\textstyle\sum_t}\kappa_{s,t}v_t \right)}
	{\int_{\RR^2}\exp\left({\textstyle\sum_t}\kappa_{s,t}v_t\right)\dd x},
\label{BLsyst}
\eeq
and \refeq{UsPDE}  into 
\beq
-\Delta V_s = 2\pi {\textstyle\sum_t}\kappa_{s,t}
   \frac{\, \exp\left( 2V_t \right)}
	{\int_{\RR^2}\exp\left(2V_t\right)\dd x},
\label{LsystemB}
\eeq
	Equations \refeq{BLsyst} and \refeq{LsystemB} are
explicit alternate representations of the Liouville system associated to
the Bennett model. 
	The coefficient matrix~$\kappa$ is manifestly non-symmetric 
in general, having negative diagonal and positive off-diagonal elements.
	Note that in the conformal case \refeq{conform}, viz.
${\textstyle\sum_t}\kappa_{s,t} =2$ for $s=\pm$, the Ansatz $v_+=v_- = v$ in
\refeq{BLsyst}, respectively $V_+=V_- = v$ in \refeq{LsystemB}, reduces both
\refeq{BLsyst} and \refeq{LsystemB} to Liouville's equation \refeq{liouEQ}.
\smallskip

\subsection{Isoperimetric estimates}
	Let $G_s$ continue to stand for either 
$G_s^{\rm TF}$ or $G_s^{\rm B}$.
	We introduce $g_s$, the primitive of $G_s$, i.e., 
$g_s^\pr(w) = G_s(w)$ for $w\in\RR$, such that the integrals 
\beq
	\int_{\RR^2} g_s(U_s)\;\dd x 
= 
	M_s 
\label{Ms}
\eeq
exist (notice that $M_s$ is defined by \refeq{Ms}).
	In each case this primitive $g_s$ is unique and given by
\beq
g_s^{\rm TF}(U_s)
= 
\beta_s^{-1}\frac{2}{h^3}\int_{\RR^3}\!
\ln\left[
 1+e^{\beta_s\! \left(
	\mu_s - c\sqrt{m_s^2c^2+|p|^2} +\nu_s cp\cdot a	+U_s\right)}\!
	\right]\!\dd p
\label{TFdensPRIM}
\eeq
for the semi-classical model, and by
\beq
g_s^{\rm B}(U_s)
= 
\beta_s^{-1}\frac{2}{h^3}\int_{\RR^3}
e^{\beta_s\! \left(
	\mu_s - c\sqrt{m_s^2c^2+|p|^2} +\nu_s cp\cdot a + U_s \right) }
\dd p
\label{BdensPRIM}
\eeq
for the classical model.
	Notice that in the classical model we have $M_s =\beta_s^{-1}N_s$, 
while in the semi-classical model we have $M_s > \beta_s^{-1}N_s$ by the 
simple convexity inequality $\ln x \leq -1 +x$, with ``$=$'' only for $x=1$.
	Notice furthermore that, in either case, the map 
$w\mapsto g_s(w)$ is monotonic increasing.
\smallskip

\noin
{\bf Lemma 4.1:} 
{\it Let the pair $(u_+,u_-)$ solve equations \refeq{usPDE} and
\refeq{usNs}, $s=\mp$, under the asymptotic conditions
\refeq{usAC}, with $\gamma$ given in \refeq{gMATRIX} satisfying 
\refeq{detMATRIX}.
	Then 
\beq 
\frac{1}{2} {\textstyle\sum_{s,t}}
\gamma_{s,t} N_sN_t - {\textstyle\sum_s} M_s \geq 0
,\label{isoperEST}
\eeq
and equality in \refeq{isoperEST} holds if and only if both $u_+$ and
$u_-$ are radially symmetric and decreasing about the same point.}
\smallskip

\noin
{\it Proof:} 
	We follow the general reasoning of 
\cite{chakieGAFA, chakieCR, chakieNA}.

	Since, by hypothesis, the pair $(u_+,u_-)$ solves the equations
\refeq{usPDE} and \refeq{usNs}, $s=\mp$, under the asymptotic 
conditions \refeq{usAC}, then $(U_+,U_-)$ satisfies  \refeq{UsPDE} 
and \refeq{usNs}, $s=\mp$, under the asymptotic conditions \refeq{UsAC}.
	Therefore, as $|x|\to\infty$,
\beq
	G_s(U_s)(x)
= 
	\frac{2}{h^3}\int_{\RR^3}
 e^{\beta_s\! \left(\mu_s - c\sqrt{m_s^2c^2+|p|^2}+\nu_s cp\cdot a\right)}
\dd p\,
{\displaystyle |x|^{-2\beta_s\sum_t\gamma_{s,t}N_t(1+\theta(x))}}
,\label{GsAC}
\eeq
with $\theta(x) = o(1)$. 
	Also by hypothesis, \refeq{UsNs} is satisfied, so that from 
\refeq{GsAC} we conclude that $\beta_s{\sum_t}\gamma_{s,t}N_t \not\!\!{<}\,1$.
	Then, by \refeq{UsAC} again, and since $U_s\in C^{2,\alpha}$ (hence, 
$U_s\in C^\infty$ by bootstrapping), the level sets 
$\Lambda^s_\xi = \{x| U_s \geq \xi_s\}$ are compact, hence
$|{\Lambda_\xi^s}| <\infty$.

	Let $x\mapsto U_s^\ast(|x|)$ 
denote the equi-measurable, radially symmetric, non-increasing 
rearrangement of $x\mapsto U_s(x)$,  centered at the origin, and 
denote by ${\Lambda^s_\xi}^\ast = \{x|\; U_s^\ast \geq \xi_s\}$ the 
ball of radius $r_\xi^s$, centered at the origin. 
	By Sard's theorem the $C^\infty$ regularity of the 
$U_s$ implies that the outward normal $\hat{\lambda}$ to $\pd\Lambda_\xi^s$ 
exists except at most for $\xi$-values in a set of measure zero, so that
the ensuing manipulations involving $\hat{\lambda}$ to $\pd\Lambda_\xi^s$ 
are well defined $\xi$-a.e.

	First, recalling that $G_s>0$, we note that on $\pd\Lambda_\xi^s$
we have $-\langle\hat{\lambda}, \nabla U_s\rangle = |\nabla U_s|$, by the Hopf lemma.
	Integration of this identity over $\pd\Lambda_\xi^s$, a trivial 
rewriting, and 
an application of the Cauchy--Schwarz inequality now gives the estimate
\beq
   -\int_{\partial\Lambda^s_\xi}
	\langle\hat{\lambda}, \nabla U_s\rangle\, \dd\sigma
=
   \int_{\partial\Lambda^s_\xi}
	| \nabla U_s|^2 \frac{1}{|\nabla U_s|}\, \dd\sigma
\geq 
   \left(\int_{\partial\Lambda^s_\xi}\, \dd\sigma\right)^2
\left(\int_{\partial\Lambda^s_\xi}\frac{1}{|\nabla U_s|}\,\dd\sigma\right)^{-1},
\label{BANDLEa}
\eeq
with equality holding if and only if $|\nabla U_s|$ is constant on 
$\partial\Lambda^s_\xi$.
	Noting that
\beq
	\int_{\partial\Lambda^s_\xi}\dd\sigma = |\Lambda^s_\xi|,
\label{BANDLEb}
\eeq
and applying the classical isoperimetric inequality~\cite{bandleBOOK}, we have
\beq
	|\Lambda^s_\xi| \geq |\partial{\Lambda_\xi^s}^\ast|,
\label{BANDLEc}
\eeq
with equality holding if and only if, up to translation,
$\partial\Lambda^s_\xi=\partial{\Lambda_\xi^s}^\ast$.
	By the co-area formula~\cite{federer}, 
\beq	
   \int_{\partial\Lambda^s_\xi}\frac{1}{|\nabla U_s|}\, \dd\sigma
= 
   \int_{\partial{\Lambda^s_\xi}^*}\frac{1}{|\nabla U_s^*|}\, \dd\sigma.
\label{BANDLEd}
\eeq 
	Pulling these estimates together we have
\beq
   -\int_{\partial\Lambda^s_\xi}
	\langle\hat{\lambda}, \nabla U_s\rangle\, \dd\sigma	
\geq
	|\partial{\Lambda_\xi^s}^\ast|^2  
\left(\int_{\partial{\Lambda^s_\xi}^*}\frac{1}{|\nabla U_s^*|}\, \dd\sigma
\right)^{-1},
\label{BANDLEe}
\eeq 
with equality holding if and only if, 
(i), $|\nabla U_s|$ is constant on $\partial\Lambda^s_\xi$, and 
(ii), $\partial\Lambda^s_\xi=\partial{\Lambda_\xi^s}^\ast$, up to translation.
	This last remark implies in particular that we can restate 
\refeq{BANDLEe} as~\cite{bandleBOOK},
\beq
   -\int_{\partial\Lambda^s_\xi}
	\langle\hat{\lambda}, \nabla U_s\rangle\, \dd\sigma	
\geq
  -\int_{\partial{\Lambda_\xi^s}^\ast}\partial_r U_s^\ast\, \dd\sigma .
\label{isoperINEQcoarea}
\eeq 

	Next, using Green's theorem and \refeq{UsPDE}, then rearrangement identity for $s=t$,
then rearrangement inequality for $s\neq t$ (in which case $t= -s$), noting that $\gamma_{s,-s} >0$
and recalling that $w\mapsto G_s(w)$ is increasing, we have
\bea
-\int_{\partial\Lambda^s_\xi}
	\langle\hat{\lambda}, \nabla U_s\rangle\, \dd\sigma 
\!\!\!\!\!\!
&&
=
	-\int_{\Lambda^s_\xi}\Delta U_s\,\dd x
\\
&& 
= 
	4\pi  {\textstyle\sum_t}\gamma_{s,t} 
	\int_{\Lambda^s_\xi}G_t({U_{t}})\,\dd x
\\
&& = 
 4\pi\Biggl(\gamma_{s,s}\int_{{\Lambda^s_\xi}^\ast}\!\!
G_s({U^\ast_{s}})\,\dd x
	+ \gamma_{s,-s}\int_{\Lambda^s_\xi}\!\!G_{-s}({U_{-s}})\,\dd x\Biggr)
\\
&& \leq 
	4\pi {\textstyle\sum_t}\gamma_{s,t} 
	\int_{{\Lambda^s_\xi}^\ast}G_t({U^\ast_{t}})\,\dd x 
\label{isoperINEQprodfunc}
\eea
where equality in \refeq{isoperINEQprodfunc}
can hold only if $U_{t}$ and $U_s$ share their  level lines (up to the labelling)
in $\Lambda_\xi^{t}$, for our $\gamma$ is irreducible.

	Combining  inequalities \refeq{isoperINEQcoarea} and \refeq{isoperINEQprodfunc}, we arrive
at the inequality 
\beq
  -\int_{\partial{\Lambda_\xi^s}^\ast}\partial_r U_s^\ast\, \dd\sigma
\leq
	4\pi {\textstyle\sum_t}\gamma_{s,t} 
	\int_{{\Lambda^s_\xi}^\ast}G_t({U^\ast_{t}})\,\dd x ,
\label{MAINineq}
\eeq
where equality can hold if and only if each $\Lambda^s_\xi$ is a disk, 
with $|\nabla U_s|$ constant on $\partial\Lambda_\xi^s$, and all the $U_s$ 
share their level lines (up to the labelling). 
	Thus, in case of equality in \refeq{MAINineq}, from the first two conditions for equality 
it follows that the family of disks $\Lambda^+_\xi$ and the family of disks $\Lambda^-_\xi$ 
are separately concentric, while from the third condition for equality it then follows that
the families of disks must be jointly concentric. 

	On the other hand, if at least one of the $U_s$ is not radially
symmetric decreasing about any point, let $\Xi^s$ be the image under $U_s$ of the 
(generally non-radial) set $\subset\RR^2$ which supports the non-radial parts of $U_s$.
	Then $\Xi^s$ has finite measure.
	Since equality in \refeq{MAINineq} cannot hold for $\xi\in \Xi^s$,  
for $\xi\in \Xi^s$ we now conclude that we have strict inequality in 
\refeq{MAINineq}, 
\beq
	-2\pi r^s_\xi U_s^{\ast\prime}(r^s_\xi) 
< 
	4\pi{\textstyle\sum_t}\gamma_{s,t} 
	\int_{{\Lambda^s_\xi}^\ast}G_{t}({U^\ast_{t}})\,\dd x
\label{isostrictINEQ}
\eeq
for {\it both} $s=\mp$.

	We now set
\beq
\cN_s(r)=\int_{B_r(0)}G_s({U_s^\ast})\,\dd x,
\eeq
and
\beq
\cM_s(r)=\int_{B_r(0)}g_s({U_s^\ast})\,\dd x
. \eeq
	We have $\lim_{r\to\infty}\cN_s(r) = N_s$
and $\lim_{r\to\infty}\cM_s(r) = M_s$, for 
\beq
\int_{\RR^2}f({U_s^\ast})\,\dd x = \int_{\RR^2}f({U_s})\,\dd x 
, \eeq
where $f$ stands for either $g_s$ or $G_s$. 
	By \refeq{isostrictINEQ}, 
\beq
2\pi r U_s^{\ast\prime}(r) \geq
-  4\pi{\textstyle\sum_t}\gamma_{s,t}\; \cN_{t}(r)
, \eeq
from which we conclude that
\beq
 r \cM_s^{\ppr}(r) \geq   \cM_s^\pr(r)-
 2 \cN^\pr_s(r){\textstyle\sum_t}\gamma_{s,t}\; \cN_{t}(r)
,\label{isoANW}
\eeq
with ``$>$'' valid for all $r>0$ for which $U_s^\ast(r) \in \Xi^s$, 
while ``$=$''  holds for $U_s^\ast(r) \not\in \Xi^s$.
	We now sum  \refeq{isoANW} w.r.t. $s=\mp$, obtaining
\beq
	{\textstyle\sum_{s}}  r \cM_s^{\ppr}(r) 
\geq  
	{\textstyle\sum_{s}} \cM_s^\pr(r)
-
	{\textstyle\sum_{s,t}}\gamma_{s,t}\Bigl(\cN_s(r)\cN_{t}(r)\Bigr)^\pr 
, \label{totDIFF}
\eeq
where we made use of the fact that $\gamma$ is real symmetric.
	Next we integrate \refeq{totDIFF} from $r=0$ to $r=\infty$,
using integration by parts on the left-hand side.
	Since $g_s({U_s^\ast})\in L^1(\RR^2)$ is radially 
decreasing, we have
\beq
\lim_{r\to\infty} r \cM_s^\pr(r) =0
, \eeq 
thus we get the result 
\beq
   \frac{1}{2}{\textstyle\sum_{s,t}}\gamma_{s,t}N_sN_t-{\textstyle\sum_s}M_s
\geq 
	0
. \label{isoINEQfinal}
\eeq

	Now, if ``$=$'' holds in \refeq{isoINEQfinal},
then all the level curves $\pd\Lambda_\xi^s$ are circles 
with $|\nabla U_s|$ constant on $\pd\Lambda_\xi^s$; 
hence~\cite{brothersziemer} the circular level curves of each $U_s$ are 
concentric, and then $U_s(x) = U_s^*(|x-x_0^s|)$ for some $x_0^s$.
	Moreover, since in case of ``$=$'' in  \refeq{isoINEQfinal} 
\refeq{isoperINEQprodfunc} tells us that the two $U_s$ must share their 
level curves (with generally different level values, of course), 
we conclude that $x_0^+ = x_0^-$, i.e. $U_+$ and $U_-$ are then 
radially symmetric and decreasing  about the same center of 
symmetry, $x_0$.

	On the other hand, if at least one of the $U_s$ is not
radially symmetric and decreasing about any point, then 
the integration picks up all the strict inequalities
from  $\xi\in \Xi^s$, and ``$>$'' holds in \refeq{isoINEQfinal}.

	Finally, since $\rank (\gamma) =2$, it follows that at
least one $u_s$ is  not radially symmetric and decreasing
about any point if at least one $U_s$ is not. 
	In the same vein,  $u_+$ and $u_-$ are radially symmetric 
and decreasing  about the same center of symmetry, $x_0$, whenever both 
$U_+$ and $U_-$ are. 

This proves Lemma 4.1. $\QED$

\subsection{Asymptotic control near infinity}
	Standard harmonic analysis gives us:
\smallskip

\noindent
{\bf Proposition 4.2:} 
{\it Under the hypothesis stated in Lemma 4.1, each solution pair 
$(u_+,u_-)$ of \refeq{usPDE},  \refeq{usNs}, \refeq{usAC}
satisfies the integral representation 
\beq  
	u_s(x) - u_s(0)
=  
	\int_{\RR^2}\left(\ln \frac{1}{|x - y|^2}-\ln\frac{1}{|y|^2}\right)
	G_s \Big({\textstyle\sum_t} \gamma_{s,t}u_t\Big)(y)\dd y
. \label{usINTrep}
\eeq
}
\smallskip

\noindent
{\bf Corollary 4.3:} 
{\it By \refeq{usINTrep} we have
\beq  
	\nabla u_s(x)
= 
	- 2\int_{\RR^2}\frac{x - y\,}{|x - y|^2}\,
	G_s \Big({\textstyle\sum_t} \gamma_{s,t}u_t\Big)(y)\dd y
. \label{GRADusINTrep}
\eeq
}
\smallskip

	With the help of \refeq{usINTrep} and \refeq{GRADusINTrep}
we obtain asymptotic control over the r.h.s. of \refeq{usPDE}, 
expressed in terms of the $U_s$. 
\smallskip

\noin
{\bf Lemma 4.4:} 
{\it Under the hypotheses of Lemma 4.1, there exists an $r_0(U_s) >0$, 
a constant $C_s>0$, and a monotonic decreasing $h_s(|x|) > 0$ satisfying
\beq
	 \lim_{|x|\to\infty} |x|^{-h_s(|x|)} =0
,\label{hDECAY}
\eeq
such that for $s=\mp$ we have, for $|x| > r_0$, 
\beq
	G_s(U_s)(x)
\leq 
	C_s |x|^{-2 -h_s(|x|)} 
.\label{GsUsASYMP}
\eeq
	Furthermore, for at least one $s$,  we have
$h_s(|x|) \geq \eps_s>0 $ for $|x| > r_0$.}
\smallskip

\noin
{\it Proof:} 
	The bound \refeq{GsUsASYMP}, with $h_s(|x|) = O(1)$ 
monotonic decreasing, follows directly from \refeq{GsAC} and \refeq{UsNs}.
	Furthermore, by \refeq{GsAC} and \refeq{UsNs} we can find $h$
such that $\int_{1}^\infty |x|^{-1-h_s(|x|)} \dd |x| <\infty$, 
but this is impossible if $ |x|^{-h_s(|x|)} \not\to 0$; hence,
\refeq{hDECAY} follows.

	This still allows $h_s(|x|) = o(1)$ for both $s$, but by 
Lemma 4.1 and the fact that $M_s\geq \beta_s^{-1}N_s$ (see the definition
of the $g_s$ above), we find, after multiplying \refeq{isoINEQfinal} 
by $-2$ and re-grouping terms,  that
\beq
	{\textstyle\sum_s} \beta_s^{-1} N_s 
	\left(2 - \beta_s {\textstyle\sum_{t}}\gamma_{s,t} N_t\right) 
\leq  
	0
. \label{isoINEQestimate}
\eeq
	Thus, if for one of the $s$  we have $h_s(|x|) = o(1)$, say for $s=+$,
then by \refeq{GsAC} we have $\beta_+ \sum_t\gamma_{+,t} N_t = 1$, so that
\refeq{isoINEQestimate} gives us right away
\beq
	\beta_{-}{\textstyle\sum_{t}}\gamma_{-,t} N_t 
\geq  
	2+ \frac{\beta_-}{\beta_+}\frac{N_+}{N_-}
. \label{isoINEQestimateB}
\eeq
	By symmetry, the analog conclusion holds for 
$\beta_{+}{\textstyle\sum_{t}}\gamma_{+,t} N_t$ if $h_-(|x|)=o(1)$.
	Hence, $h_s(|x|) = o(1)$ for at most one of the $s$. 

	This proves Lemma 4.4. $\QED$
\smallskip

\noin
{\bf Corollary 4.5:} 
{\it For at  least one $s$, we have
$\left|\int_{\RR^2}\ln |y|\; G_s \big(U_s\big)(y)\dd y\right|<\infty$, 
so that for this $s$, we have
\beq
	\lim_{|x|\to\infty}\Big(u_s(x) + 2N_s\ln|x|\Big) 
=
	u_s(0) + 2 \int_{\RR^2}\ln |y| G_s \big(U_s\big)(y)\dd y
.\eeq
}
\smallskip

	We proceed with gradient estimates. 
\smallskip

\noin
{\bf Lemma 4.6:} 
{\it Under the hypothesis stated in Lemma 4.1, each solution pair 
$(u_+,u_-)$ of \refeq{usPDE}, \refeq{usNs}, \refeq{usAC} 
satisfies the gradient estimates
\beq
	\limsup_{|x|\to\infty}|x||\nabla u_s| 
\leq 
	2N_s
.\label{modGRADuEST}
\eeq
}
\smallskip

\noin
{\it Proof:}
	By Corollary 4.3, we have
\beq
	\big|\nabla u_s(x)\big| 
\leq 
	2\int_{\RR^2}{\frac{G_s\!\left(U_s\right)(y)}{|x-y|}}\, \dd y
.\label{modGRADuESTa}
\eeq
	After multiplying \refeq{modGRADuESTa} by $|x|$, a simple rewriting
of the r.h.s. gives
\beq
	|x|\big|\nabla u_s(x)\big|
\leq 
	2 \int_{\RR^2}G_s(U_s)(y)\dd y 
	+ 2 \int_{\RR^2}\left({\frac{|x|}{|x-y|}}-1\right)G_s(U_s)(y)\dd y 
.\label{modGRADuESTb}
\eeq
	By \refeq{UsNs} the first integral on the r.h.s. of 
\refeq{modGRADuESTb} equals $N_s$.
	By the triangle inequality, the second integral on the
r.h.s. of \refeq{modGRADuESTb} is bounded in absolute value by
\beq
2\int_{\RR^2}{\frac{|y|}{|x-y|}}G_s(U_s)(y) \dd y 
.\label{modGRADuESTc}
\eeq
	We now show that 
\beq
\lim_{|x|\to\infty} \int_{\RR^2}{\frac{|y|}{|x-y|}}G_s(U_s)(y)\dd y=0
,\label{modGRADuESTd}
\eeq
from which the lemma follows.

	We split the domain of integration in \refeq{modGRADuESTc} as follows:
\hbox{$\RR^2 = \Omega_1{\cup}\Omega_2{\cup}\Omega_3$,} with
$\Omega_1 = \{ y\mid|y|<|x|/2 \}$, 
$\Omega_2 =\{y\mid|x|/2\leq|y|\leq 2|x|\}$, and
$\Omega_3 = \{y\mid|y|>2|x|\}$. 
	If $G_s(U_s)(y) \leq C |y|^{-2-\eps}$, 
with $0<\epsilon <1$, then the estimates are precisely the same as 
in \cite{chakieGAFA}, section 2, with $\exp$ replaced by $G_s$; this 
is the case for at least one of the $s$. 
	It remains to provide estimates when $h_s(|x|) = o(1)$
for  one of the $s$. 

	To estimate the contribution from $\Omega_1$ when 
$G_s(U_s)(y) \leq C |y|^{-2-h_s(|y|)}$ with $h_s(|y|) = o(1)$,
we note that
\bea
	\int_{\Omega_1}\!{\frac{|y|}{|x{-}y|}} G_s(U_s)(y) \dd y
\leq
	\frac{C}{|x|}\int_{\Omega_1}\!|y|G_s(U_s)(y) \dd y
\leq 
	\frac{C'}{|x|}\int_{0}^{|x|} \zeta^{-h_s(\zeta)} \dd \zeta
.\label{modGRADuESTe} 
\eea
	As for the \rhs of \refeq{modGRADuESTe}, L'Hopital's Rule  gives us
\beq
  \lim_{|x|\to\infty}\frac{C'}{|x|}\int_{0}^{|x|}\zeta^{-h_s(\zeta)} \dd \zeta
=
   C' \lim_{|x|\to\infty} |x|^{-h_s(|x|)} =0
,\label{modGRADuESTeB} 
\eeq
the last step by Lemma 4.4. 
	Hence, the l.h.s.\refeq{modGRADuESTe} vanishes  as $|x|\to\infty$.

	Similarly, the contribution from $\Omega_2$ is 
estimated by using again that $G_s(U_s)(y) \leq C |y|^{-2-h_s(|y|)}$,
so that for $|x|$ large enough we have the  bound
\beq
	\int_{\Omega_2}{\frac{|y|}{|x-y|}}G_s(U_s)(y)\dd y 
\leq
	\frac{C}{|x|^{1+h_s(|y|)}}\int_{|y|<4|x|}{\frac{\dd y}{|y|}}
\leq
	C|x|^{-h_s(|x|)}
.\label{modGRADuESTg}
\eeq
	Clearly r.h.s.\refeq{modGRADuESTg}$\to 0$  as $|x|\to\infty$,
by the same reasoning as for $\Omega _1$. 

	Finally, the contribution from $\Omega_3$ is dominated by
\beq
	\int_{\Omega_3}{\frac{|y|}{|x-y|}}G_s(U_s)(y) \dd y 
\leq
	C\int_{|y|>2|x|}G_s(U_s)(y) \dd y
,\label{modGRADuESTf} 
\eeq
which vanishes as $|x|\to\infty$, by hypothesis \refeq{UsNs}.

	This concludes the proof of Lemma 4.6. $\QED$
\smallskip

\noin
{\bf Lemma 4.7:} 
{\it Under the hypotheses of Lemma 4.1, we have, uniformly in $x$,
\beq
	\lim_{|x|\to\infty}\langle x,\nabla u_s\rangle = - 2 N_s.
\label{LEMMAfourSEVEN}\eeq
}
\smallskip

\noin
{\it Proof:} 
	Let $\hat{x}={x/{|x|}}$ and $\hat{y}={y/{|y|}}$, with $|x|=|y|$.  
	Now fix $\hat{x}\in \SS^1$.  
	By \refeq{usAC}, we have
\beq
	\lim_{\tau\to\infty}\frac{u_s(\tau\hat{x})}{\ln \tau}
=
	- 2N_s
\eeq
	Thus, by L'Hopital's Rule,
\beq
	\lim_{\tau\to\infty}\tau\frac{\dd}{\dd \tau}u_s(\tau\hat{x})
=
	\lim_{|x|\to\infty} \langle x,\nabla u_s\rangle
=
	-2N_s
\label{lhospital}
\eeq
for $x =|x|\hat{x}$.
	It remains to establish uniformity of \refeq{lhospital}. 
	To this extent, we show that there exist $R$ and $\delta$, 
such that, if $|x|>R$ and $|\hat{x}-\hat{y}|<\delta$, then
\beq
 \big|\langle x,\nabla u_s(x)\rangle - \langle y,\nabla u_s(y)\rangle\big| 
< 
\eps 
.\label{uniformGRAD}
\eeq

	We first show that for $|x|>R$ and $|x-y|<{|x|/ {10}}$, we have,
\beq
	|x|\big|\nabla u_s(x)-\nabla u_s(y)\big|
\leq 
	C|\hat{x}-\hat{y}|+ C' |x|^{-h_s(|x|)} .
\eeq
	By Corollary 4.3, 
\beq
	\big|\nabla u_s(x)-\nabla u_s(y)\big|
\leq 
	2\int_{\RR^2}G_s(U_s)(z)\left|{\frac{x-z}{|x-z|^2}}
	-{\frac{y-z}{|y-z|^2}}\right|\dd z 
.\eeq
	We break up the domain of integration in the above integral 
exactly as in the proof of  Lemma 4.6. 
	(Notice the integration variable is now $z$.)
	The integration over $\Omega_1$ is estimated exactly as in 
section 2 of \cite{chakieGAFA} to get
\beq
	\int_{\RR^2}G_s(U_s)(z){\frac{|x-y|}{|x|^2}}\dd z
\leq 
	C{\frac{|x-y|}{|x|^2}} 
.\eeq
	The  integral over $\Omega_2$ is dominated by
\beq
	\int_{|z|\sim |x|}G_s(U_s)(z)\left(\frac{1}{|x-z|}
	+\frac{1}{|y-z|}\right)\dd z
\leq 
	C |x|^{-1-h_s(|x|)}
.\eeq
	The final estimate above was identical to that made in the
proof of Lemma~4.6. 
	Use was made of $G_s(U_s)(z)\leq C|x|^{-2- h_s(|x|)}$ on $\Omega_2$,
which holds by Lemma 4.4.
	The contribution from $\Omega_3$ is estimated once again exactly as 
in section 2 of \cite{chakieGAFA} to be dominated by
\beq
	C|x-y|\int_{|z|>2|x|}{\frac{G_s(U_s)(z)}{|z|^2}}\,\dd z
\leq 
	C' {\frac{|x-y|} {|x|^2}}\int_{\RR^2} G_s(U_s)\dd x     
\leq 
	C^\ppr {\frac{|x-y|}{|x|^2}} 
,\eeq
where the last step follows by \refeq{UsNs}.
	By these estimates,
\bea
   \big|\langle x,\nabla u_s(x)\rangle-\langle y,\nabla u_s(y)\rangle\big| 
\!\!\!\!\!\!\!\!\!
&&
\leq 
	|x||\hat{x}-\hat{y}|\big|\nabla u_s(x)\big|
	+ |y|\big|\nabla u_s(x)-\nabla u_s(y)\big|\qquad
\\
&& 
\leq 
	|x|\big|\nabla u_s(x)\big||\hat{x}-\hat{y}|+|\hat{x}-\hat{y}|+
		C|x|^{-h_s(|x|)}
.\quad
\eea
	By Lemma 4.6, the last expression above is at most 
$C' \delta + C|x|^{-h_s(|x|)}$. 
	Thus our claim \refeq{uniformGRAD} follows now from Lemma 4.4
for suitably large $R$ and small $\delta$. 
	Since $\SS^1$ is compact, uniformity of the limit in Lemma 4.7 
now follows. $\QED$
\smallskip

	Lemmata 4.6 and 4.7 imply
\smallskip

\noin
{\bf Corollary 4.8:} 
{\it Under the hypotheses expressed in Lemma 4.1, we have, 
uniformly in $x$,
\beq
	\lim_{|x|\to\infty}|x||\nabla u_s| 
= 
	2 N_s
.\eeq
}
\smallskip

\noin
{\it Proof:} 
	Follows essentially verbatim \cite{chakieGAFA}, proof of 
Corollary 2.2, with $\exp$ replaced by $G_s$. $\QED$
\smallskip

	Let $\Omega_\xi=\{x|\; u_s(x)\geq \xi\}$, where $\xi\ll -1.$ 
	By  \refeq{usAC} it follows that if $x\in\partial \Omega_\xi$, 
then $|x|\geq R(c)$ with $R(c)$ large.
	For such $x$, it follows from Corollary 4.8 that $\nabla u_s\neq 0$. 
	Since $u\in C^{2,\alpha}_{loc}$ we easily see that therefore
$\partial\Omega_\xi\in C^{2,\alpha}$. 
	Thus the unit outward normal $\hat{\omega} (x)$ 
to $\partial \Omega_\xi$ exists at all $x\in\partial\Omega_\xi$ 
for $\xi$ sufficiently negative. 

\noin
{\bf Lemma 4.9.} 
{\it Let $\hat{\omega} (x)$ be the unit outward normal to
$\partial\Omega_\xi$ at $x$, and let $\hat{x} = x/|x|$. 
	We have, uniformly in $x$,
\beq
\lim_{c\to -\infty}\langle \hat{x},\hat{\omega}\rangle = 1 
.\eeq
}
\smallskip

\noin
{\it Proof:} 
	Identical to \cite{chakieGAFA}, proof of Lemma 2.8. $\QED$
\smallskip

\noin
{\it Remark:}
Lemma 4.9 implies that  asymptotically for large $x$
 the $\partial\Omega_\xi$ become concentric circles.
\smallskip

	We are now in a position to prove our Theorem 3.1.

\subsection{Proof of the Virial Theorem}
\noin
{\bf Proposition 4.10:} 
(Rellich--Pokhozaev identity.)
{\it Under the hypotheses expressed in Theorem 3.1, any solution pair
$(u_+,u_-)$, of \refeq{usPDE}, \refeq{usNs}, \refeq{usAC} satisfies 
the Rellich--Pokhozaev identity
\beq
	\frac{1}{2}{\textstyle{\sum_{s,t}}}\gamma_{s,t} N_s N_t 
	- {\textstyle\sum_s}   M_s 
= 
	0
. \label{relpohIDliouEQsyst}
\eeq}
\smallskip

\noin
{\it Remark:} 
	The Rellich--Pokhozaev identity \refeq{relpohIDliouEQsyst} 
is identical to the identity expressed in the Virial Theorem 3.1. 
\smallskip

\noin
{\it Proof of Proposition 4.10:} 
For $(u_+,u_-)$ a solution pair of \refeq{usPDE}, \refeq{usNs}, \refeq{usAC},
we have the partial differential identity 
\beq
{\rm div} ( \langle x,\nabla u_{t}\rangle \nabla u_s ) 
=
\langle \nabla u_s , (1 + \langle x,\nabla \rangle ) 
\nabla u_{t}\rangle
-4\pi \langle x,\nabla u_{t}\rangle G_s ({\textstyle\sum_{t}}\gamma_{s,t}u_t)
.\label{relpohANSliouEQsyst}
\eeq
	We will multiply \refeq{relpohANSliouEQsyst} by $\gamma_{s,t}$, 
sum  over $s$ and $t$, integrate over $B_R$, use some partial integrations,
then take the limit $R\to\infty$.

	We evaluate first the left-hand side of \refeq{relpohANSliouEQsyst}.
	Green's theorem gives us
\beq
\int_{B_R}
{\rm div} ( \langle x,\nabla u_{t}\rangle \nabla u_s )\dd x
=
\int_{\partial B_R}
	|x|^{-1} \langle x,\nabla u_{t}\rangle
	\langle x, \nabla u_s\rangle\, \dd\sigma
.\label{partINTrelpohSYST}
\eeq
	Taking the limit $R\to \infty$, using \refeq{LEMMAfourSEVEN}, 
we get
\beq
\lim_{R\to\infty}
\int_{\partial B_R}
|x|^{-1} \langle x,\nabla u_{t}\rangle
\langle x, \nabla u_s\rangle\, \dd\sigma
= 8 \pi N_sN_t
,\label{firstLIMIT}
\eeq
hence
\beq
\lim_{R\to\infty}
{\textstyle{\sum_{s,t}}}\gamma_{s,t}
\int_{B_R}
{\rm div} ( \langle x,\nabla u_{t}\rangle \nabla u_s )\dd x
=8\pi {\textstyle\sum_{s,t}} \gamma_{s,t} N_s N_t
.\eeq
	
	On the other hand, the last term in the right-hand side of
\refeq{relpohANSliouEQsyst} gives us
\bea
&&
\!\!\!\!\!\! \!\!\! 
{\textstyle{\sum_{s,t}}}\gamma_{s,t}\!
\int_{B_R}\!\! \langle x,\nabla u_{t}\rangle 
	G_s({\textstyle\sum_{t}}\gamma_{s,t}u_t)\dd x
 = 
{\textstyle{\sum_s}} 
\int_{B_R}\!\!
 \langle x,\nabla g_s({\textstyle\sum_{t}}
\gamma_{s,t}u_{t})\rangle\dd x
\nonumber \\ &&
= 
{\textstyle{\sum_s}} 
\int_{\partial B_R}\! |x|\;g_s({\textstyle\sum_{t}}\gamma_{s,t}u_{t})\dd\sigma
-2{\textstyle{\sum_s}}\int_{B_R}\!\!g_s({\textstyle\sum_t}\gamma_{s,t}u_{t})
\dd x\,
.\label{lastterm}
\eea
	We now take the limit $R\to \infty$ in \refeq{lastterm}.
	As for the surface integrals, we note that by Lemma 4.4 we
have $g_s(U_s)(x)\sim C G_s(U_s)(x)$ as $|x|\to\infty$, so that
once again by Lemma 4.4, we have
\beq
\lim_{R\to\infty}
 \int_{\partial B_R}
 | x |\; g_s({\textstyle\sum_t} \gamma_{s,t}u_{t}) \,\dd\sigma = 0
.\eeq
	As for the volume integrals, we get
\beq
\lim_{R\to\infty} 
 \int_{B_R}g_s({\textstyle\sum_t} \gamma_{s,t} u_{t})\,\dd x = M_s
.\eeq
	Turning now to the first term in the right-hand side of
\refeq{relpohANSliouEQsyst}, we use the symmetry of $\gamma$,  
an integration by  parts and \refeq{usPDE}, to get 
\bea
&& 
\!\!\!\!\!\!\!\!\!\!\!
{\textstyle\sum_{s,t}}\gamma_{s,t}
\int_{B_R}
\langle \nabla u_s , (1\! +\! \langle x,\nabla \rangle ) 
\nabla u_{t}\rangle \,\dd x
 =
{1\over 2}{\textstyle\sum_{s,t}}\gamma_{s,t}
\int_{\partial B_R} 
| x | \langle \nabla u_s, \nabla u_{t}
\rangle \,\dd\sigma\qquad
\\
&&
=
{1\over 2}
{\textstyle\sum_{s,t}}\gamma_{s,t}
\int_{\partial B_R} \Big(|x|^{-1} 
\langle x, \nabla u_s\rangle
		      \langle x, \nabla u_{t} \rangle 
+ | x | \langle \nabla_{\rm T} u_s, \nabla_{\rm T} u_{t}\rangle\Big) \dd\sigma
\label{alltogether}
\eea
where $\nabla_{\rm T}$ denotes tangential derivative. 
	By Lemma 4.7 and Corollary 4.8, we have
\beq
|x|^{2}|\nabla_{\rm T}u_s|^2 = 
|x|^{2}|\nabla u_s|^2 - \langle x, \nabla u_s\rangle^2 \to 0
,\eeq
uniformly as $|x|\to \infty$. 
	Thus as $R\to \infty$, 
\beq
\int_{\partial B_R} 
| x | \langle \nabla_{\rm T} u_s, \nabla_{\rm T} u_{t}\rangle \,\dd\sigma \to 0
\eeq
and therefore 
\beq
\lim_{R\to\infty} 
{\textstyle\sum_{s,t}}\gamma_{s,t}
\int_{B_R}
\langle \nabla u_s , (1 + \langle x,\nabla \rangle ) 
\nabla u_{t}\rangle \,\dd x
=
4\pi{\textstyle\sum_{s,t}}\gamma_{s,t}N_s N_t
.\label{therefore}
\eeq

	Pulling all limit results together, we obtain Proposition
4.10. $\QED$
\smallskip

\noin
{\it Remark.} The proof of the virial theorem extends to situations
when $\gamma$ does not have full rank, hence to more-than-two species beams.
\smallskip

\subsection{Concluding the proof of the Symmetry Theorem}
	By Lemma 4.1, and by Proposition 4.10, the solutions 
$u_s$ of \refeq{usPDE}, \refeq{usNs}, \refeq{usAC},
have to be radially symmetric and decreasing about a common center $x_0$.
	Since the coupling matrix $\gamma$ is invertible, the
same conclusion holds for the solutions $U_s$ of
\refeq{UsPDE}, \refeq{UsNs}, \refeq{UsAC}.
	The proof is complete. $\QED$
\medskip

\noin
{\bf Acknowledgement} The results reported here spun out of 
my collaboration with Sagun Chanillo. Work supported by
NSF Grant DMS 9623220.

%
\end{document}